\documentclass[preprint]{revtex4}
\usepackage{graphicx}
\usepackage{dcolumn}
\usepackage{bm}
\usepackage{hyperref}
\newcommand{\footremember}[2]{%
   \footnote{#2}
    \newcounter{#1}
    \setcounter{#1}{\value{footnote}}%
}
 
\begin{document}
\setcounter{page}{1}
\title
{Deuteron properties from muonic atom spectroscopy}
\author {%
N. G. Kelkar \footremember{*}{Email: nkelkar@uniandes.edu.co}%
and D. Bedoya Fierro} 
\affiliation
{Dept. de Fisica, Universidad de los Andes,
Cra.1E No.18A-10, Santafe de Bogot\'a, Colombia
}
\begin{abstract}
Leading order ($\alpha^4$) finite size corrections in muonic deuterium are evaluated 
within a few body formalism for the $\mu^- p n$ system 
in muonic deuterium and found to be sensitive to the input of the deuteron 
wave function. We show that this sensitivity, taken along with the precise 
deuteron charge radius determined from muonic atom spectroscopy can be used to 
determine the elusive deuteron D-state probability, $P_D$, for a given model 
of the nucleon-nucleon (NN) potential. The radius calculated with a 
$P_D$ of 4.3\% in the chiral NN models and about  
5.7\% in the high precision NN potentials is favoured most 
by the $\mu^-d$ data.
\end{abstract}
\maketitle


\section{Introduction} 
The lightest nucleus, namely, the deuteron, has 
traditionally held an important place in nuclear physics as a testing ground for the 
nucleon-nucleon interaction. Determining the D-state probability in 
the deuteron wave function in particular has been a classic problem of 
nuclear physics \cite{feshschwin,rodningsboth,oller}. 
Stating the problem in simple words, the deuteron has a quadrupole 
moment and hence cannot be in a pure S-state but rather a D-state admixture is required. 
However, as it was shown in \cite{amadofriar} that the 
D-state probability, $P_D = \int_0^\infty w^2(r) dr$ (with $w(r)$ being the 
deuteron radial wave function with $l =2$), is inaccessible directly to 
experiments, it is usually the asymptotic 
D-state to S-state wave function ratio, $\eta$ \cite{rodningsboth, conzet}, which is 
determined. There do exist attempts to determine $P_D$ from the measured 
magnetic moment of the deuteron, $\mu_D$, with, 
$\mu_D = \mu_S - (3/2) P_D (\mu_S 
-1/2) + \delta_R$, where, $\mu_S = \mu_P + \mu_N$ is the isoscalar nucleon 
magnetic moment. However, the term $\delta_R$ which includes  
mesonic exchange effects,
relativistic corrections, dynamical effects and
isobar configurations in the deuteron introduces uncertainties in 
the extraction of $P_D$ \cite{michael}. This fact was noticed in  
one of the oldest works by Feshbach and Schwinger \cite{feshschwin} on the 
theory of nuclear forces which 
gave the D-state probability, $P_D$, ranging between 
2\% to 6\%. Much later, Ref. \cite{mathel} listed values of 
$P_D$ ranging from 0.28 to 6.47\% for 9 different nucleon-nucleon (NN) 
potentials. However, 
earlier in \cite{levin} the possible minimum was shown to be 0.45\%. 
With $P_D$ not being a measurable quantity, 
Refs \cite{rodningsboth} and \cite{conzet} determined the asymptotic ratio 
$\eta$ = 0.0256 $\pm$ 0.0004 and 0.0268 $\pm$ 0.0013 from tensor analyzing powers 
in sub-Coulomb ($d,p$) reactions and $dp$ elastic scattering respectively. 
In the absence of a ``measured" D-state probability, 
theoretical models of the NN interaction also try 
to reproduce the asymptotic ratio $\eta$ determined from experiments 
(in addition to other data) to confirm the reliability of the NN 
model \cite{oller}.  

The purpose of this work is to present a new method which provides a 
means to fix the percentage of the ``elusive" \cite{amadofriar} 
D-state probability, $P_D$, from experiments in an indirect manner. 
The method is particularly useful in view of the very high precision 
reported by recent muonic deuterium experiments \cite{pohldeut}. 
It is based on a few body calculation of 
the leading order ($\alpha^4$)
finite size corrections (FSC) to the energy 
levels of muonic deuterium atoms. There exists extensive literature on 
corrections including the deuteron polarization 
\cite{leidman, pachucki,bacca}, with 
detailed calculations of FSC at higher orders 
($\alpha^5$, $\alpha^6$ etc) \cite{friarseminal,leidman,pachucki,bacca}.
The sensitivity of the higher order FSC to the 
form of the nucleon-nucleon potential (and hence the deuteron wave function) 
is found to be small \cite{leidman,bacca} or negligible \cite{friarzero}. 
The leading FSC at order $\alpha^4$ in these works is written in terms of 
the deuteron charge radius. 
The few body formalism of the present work helps in revealing 
the dependence of the leading FSC term on the proton and neutron form factors as well as 
the deuteron wave function. 
We show that a comparsion of the order $\alpha^4$ FSC with those 
of Ref. \cite{pohldeut} where the radius is 
precisely extracted from 
measurements in muonic deuterium provides a method to adjust the 
deuteron D-state probability. To be specific, we present calculations using different  
parametrizations of the deuteron wave function 
(with different amounts of the D-state probabilities) 
and compare the corrections with those given in \cite{pohldeut} in a form dependent on 
the deuteron charge radius, $r_d$. 
Though the general trend of the results is an increase in the radius for smaller 
values of $P_D$, the results are found to depend on the type of model used. 
In the class of chiral models \cite{chiralNN}, $P_D$ = 4.3\% is found to be 
favourable for the 
closest agreement with the precise value of $r_d = 2.12562(78)$ fm \cite{pohldeut}. 
Using high precision NN potentials such as Nijmegen, Reid, Paris etc 
\cite{highNN}, 
$P_D$ = 5.7\% to 5.8\% is favoured by the $\mu d$ data. 

\section{Finite size effects in muonic deuterium} 
Finite size corrections (FSC) to the energy levels in the hydrogen atom has been a 
topic of revived interest \cite{weproton} 
in the past few years due to the increase in the 
precision achieved in atomic spectroscopy measurements. These effects are 
manifested more strongly in muonic atoms due to the fact that the muon is about 
200 times heavier than the electron and hence has a Bohr radius which is much smaller. 
In view of the recent precise measurement of the Lamb shift in muonic deuterium 
\cite{pohldeut}, 
it seems timely to put forth the question as to what other impact (apart from 
the precise radius determination) does this measurement have on physics. 
In order to see this, we study the effects of deuteron structure on the 
energy levels in this atom. The present work considers the effects at leading order 
($\alpha^4$) and we refer the reader to \cite{leidman,pachucki,bacca} for higher 
order corrections. 

\subsection{Electromagnetic muon-deuteron potential} 
We investigate the finite size effects by calculating 
the energy correction, $\Delta E$, using first order perturbation theory involving 
an electromagnetic muon-deuteron potential, $V_{\mu^- d}$. 
The latter is constructed using a three body approach to the muon-proton-neutron 
system with the proton and neutron being bound inside the deuteron. As we will see 
below, the $\mu^- p$ and $\mu^- n$ interactions are obtained using the proton and 
neutron electromagnetic form factors and the $p n$ interaction is contained in the 
deuteron wave function.  
Such a potential 
can be constructed using standard techniques from scattering theory where we first 
write down the scattering amplitude to 
obtain the potential $V_{\mu^- d}(\bm{q})$ in momentum space and then evaluate its 
Fourier transform which enters the energy correction given by, 
$\Delta E = \int_0^{\infty}\Delta {V}(r) |\Psi_{nl}(\bm{r})|^2 d^3r$.
This procedure of obtaining potentials in coordinate space is also 
common in quantum field theory \cite{wejphysg,casimir,nuclear}. 
Here, $\Delta {V}$ is the difference of $V_{\mu^- d}(r)$ and the $\mu^- d$ 
electromagnetic 
potential assuming the deuteron to be point-like. Details of the few body formalism 
used here can be found in \cite{belyaev, myprl}. 
We shall repeat the relevant steps briefly below. 

The Hamiltonian of the quantum system consisting of a muon and a nucleus 
(with A nucleons) is given as \cite{belyaev},  
$H = H_0 \, + \, V_{\mu^- A}\, +\, H_A$, 
where $H_0$ is the muon-nucleus kinetic energy operator (free Hamiltonian), 
$V_{\mu^- A} = \sum_{i=1}^A\, V_i $, the sum of muon-nucleon potentials, 
$V_i \equiv V_{\mu^- N}(|\bm{R} - \bm{r}_i|)$, where $\bm{R}$ and $\bm{r}_i$ 
are the coordinates of the muon and the $i^{th}$ nucleon with respect to the centre 
of mass of the nucleus and 
$H_A$ is the total Hamiltonian of the nucleus containing the potential term, 
$\sum_{i \ne j}\, V_{NN} (|\bm{r}_i - \bm{r}_j|)$. 
We proceed with the assumption that the nucleus remains in its ground 
state during the scattering process, i.e., 
$H_A \, | \Phi \rangle = \epsilon \, | \Phi \rangle$ and that the nucleons occupy 
fixed positions inside the nucleus. 
The muon - nucleus elastic scattering amplitude can be expressed as \cite{belyaev} 
$f (\bm{k}^{\prime}, \bm{k}; E) = - (\mu/\pi)\, \langle\, \bm{k}^{\prime}, 
\Phi \, |\,T(E)\,|\bm{k}, \Phi\, \rangle$ in terms of the matrix elements of the 
operator $T$ obeying the Lippmann-Schwinger (L-S) equation, 
$T = V + V  (E - H_0 - H_A)^{-1} T$. $|\bm{k}, \Phi \,\rangle$ and 
$|\bm{k}^{\prime}, \Phi \rangle$ are the initial and final asymptotic states 
which differ only in the direction of the relative muon nucleus momenta 
$\bm{k}$ and $\bm{k}^{\prime}$.  
Since the electromagnetic potential, $V_{\mu^- \,A}$,  is 
proportional to the coupling constant 
$\alpha \sim 1/137$, it is reasonable to truncate the L-S equation 
at first order and approximate $T = V = \sum_i V_i$. 
Thus, $T(\bm{k}^{\prime},\bm{k})$ = 
$V(\bm{k}^{\prime},\bm{k})$ and denoting, $T(\bm{k}^{\prime},\bm{k}) 
 \equiv \langle\, \bm{k}^{\prime}, 
\Phi \, |\,T(E)\,|\bm{k}, \Phi\, \rangle$, we have 
$V(\bm{k}^{\prime},\bm{k})\,=\, \langle \, \bm{k}^{\prime}, 
\Phi \, |\,\sum_{i=1}^A \,V_i\,|\bm{k}, \Phi\, \rangle $. If the internal Jacobi 
coordinates are denoted by $\bm{x}_i$, then relating them with 
$\bm{r}_i = a_i \bm{x}_1 \,+\,b_i \bm{x}_2 \,+\, ...\,+\,g_i \bm{x}_{A-1}$, we 
can write, 
\begin{equation}\label{pot1}
V(\bm{k}^{\prime},\bm{k}) = \int d\bm{x}_1\, d\bm{x}_2\, ...
\,d\bm{x}_{A-1}\, |\Phi(\bm{x}_1,\bm{x}_2,\, ...)|^2 \sum_{i=1}^A 
\, V_i (\bm{k}^{\prime}, \bm{k}, \bm{r}_i)\, , 
\end{equation}
where, $V_i (\bm{k}^{\prime}, \bm{k}, \bm{r}_i) = V_i(\bm{k}^{\prime}, \bm{k})\, 
\exp [i (\bm{k} - \bm{k}^{\prime}) \cdot \bm{r}_i ]$. The above discussion is valid 
for any nucleus with $A$ nucleons. In case of the muon-deuteron system, this reduces to 
\begin{equation}\label{potdeut}
V(\bm{k}^{\prime},\bm{k}) \, =\, \int\, d\bm{x}_1\,|\Phi_d(\bm{x}_1)|^2 
\, [\, V_{\mu^- p} (\bm{k}^{\prime}, \bm{k}, {1\over 2}\bm{x}_1)\, + 
\, V_{\mu^- n} (\bm{k}^{\prime}, \bm{k}, -{1\over 2}\bm{x}_1)\,]\, 
\end{equation}
where we used, $\bm{x}_1 = \bm{r}_1 - \bm{r}_2$, $\bm{r}_1 = (1/2) \bm{x}_1$ and 
$\bm{r}_2 = -(1/2) \bm{x}_1$. 
We identify $1$ and $2$ with
proton and neutron so that, $V_1 = V_{\mu^- p}$, $V_2 = V_{\mu^- n}$  
and $\Phi_d$ is the deuteron wave function. 

To evaluate (\ref{potdeut}), we need the $\mu^-$-nucleon 
electromagnetic potential, which, with the inclusion of 
the nucleon electromagnetic form factors $G_E^N(q^2)$ can be 
written using the formalism of the Breit equation \cite{wejphysg} within the 
one-photon-exchange interaction. Since 
such a potential was explicitly derived in \cite{weproton, wejphysg} 
by evaluating the elastic muon-nucleon amplitude expanded in powers of $1/c^2$, 
we shall not repeat the derivation here. 
This potential with form factors contains  
23 terms \cite{wejphysg} corresponding to the (i) Coulomb potential, 
(ii) Darwin terms, and (iii) 
spin dependent terms which give rise to fine and hyperfine structure.
If we consider only the scalar parts of the Breit potential, they depend only 
on $\bm{q}^2$ and hence we can write, 
$V_{\mu^- N}(\bm{k},\bm{k}^{\prime}) \equiv V_{\mu^- N}(\bm{q})$ 
\cite{wejphysg,weproton}, 
where, $\bm{q} = \bm{k} - \bm{k}^{\prime}$ is the momentum transfer carried by the 
exchanged photon. 
Denoting $Q = |\bm{q}|$, the $\mu^-N$ potential is 
given as \cite{weproton},
\begin{equation}\label{potmuN}
V_{\mu^- N}(Q) =  -4 \pi \alpha {G_E^N(Q^{2})\over Q^{2}} \, 
\biggl \{ 
1 - {Q^2 \over 8 m_N^2 c^2} - {Q^2 \over 8 m_{\mu}^2 c^2} \biggr \}\, , 
\end{equation}
where $m_N$ and $m_{\mu}$ are the nucleon and muon masses. $G_E^N(Q^2)$ is the 
nucleon electric form factor. 
A Fourier transform of the first term in 
the curly bracket leads to the $\mu^- N$ Coulomb potential for a finite 
sized nucleon. The next two terms in the curly brackets are 
relativistic corrections, 
the Darwin terms in the muon (spin 1/2) - 
nucleon (spin 1/2) $\mu^-N$ interaction Breit potential. 
The Darwin term ${Q^2/ 8 m_N^2 c^2}$
is conventionally not considered as a part of the nucleon  
form factor $G_E^N(q^2)$ \cite{friaradv} and hence is 
kept explicitly in the muon-nucleon potential here. 
Putting together (\ref{potdeut}) and (\ref{potmuN}) we obtain 
the muon-deuteron electromagnetic potential, 
$
V_{\mu^- d} (Q) = V_{\mu^- p}(Q) \int \,d\bm{x} \,|\Phi_d(\bm{x})|^2 
\, e^{-i \bm{q} \cdot \bm{x}/2} \, +\, V_{\mu^- n}(Q) \int\, d\bm{x} \, 
|\Phi_d(\bm{x})|^2 
\, e^{i \bm{q} \cdot \bm{x}/2}$, in momentum space. 
The integrals in this expression can be shown to reduce to \cite{mathel} 
$G_0(Q) = \int_0^{\infty} \, [ u^2(r) + w^2(r)] \, j_0(Qr/2) \, dr$, where, $u(r)$ and 
$w(r)$ are the radial parts of the deuteron S- and D-wave functions. Thus, 
$V_{\mu^-d}(Q) = (V_{\mu^- p}(Q) + V_{\mu^- n}(Q)) G_0(Q)$, so that, 
\begin{equation}\label{potqspace}
V_{\mu^-d}(Q) = - 4 \pi \alpha\, 
{G_0(Q) [ G_E^p(Q^2) + G_E^n(Q^2)] \over Q^2} \, \biggl ( 1 \, -\, 
{Q^2 \over 8 m_N^2} \, -\, {Q^2 \over 8 m_{\mu}^2} \biggr )\, , 
\end{equation}
where the proton and neutron masses have been written as $m_p \approx m_n \approx m_N$
for simplicity. 
We note here that the three body formalism allows us to include the 
relativistic corrections in the form of the Darwin terms since we are summing potentials 
between the muon and nucleons (both of which are spin - 1/2 objects) and folding 
them with the nuclear structure part. Including the relativistic corrections 
directly in a muon-deuteron potential is otherwise a formidable task since one has 
to work with an equation for spin 1/2 - spin 1 elastic scattering with form factors. 
The above Darwin term is known as a recoil correction in 
atomic physics (see \cite{krauth} for a detailed discussion). 

The elementary potential (\ref{potmuN}) is calculated using the dipole 
proton form factor, $G_E^p(Q^2) = (1 + Q^2/0.71)^{-2}$ 
and a Galster form for the neutron, 
$G_E^n (Q^2) = [1.91\tau/(1 + 5.6 \tau)] (1 + Q^2/0.71)^{-2} $ 
(with $\tau = Q^2/4 m_n^2$) as in \cite{gilmangross}. 
These particular forms were chosen since using these forms of $G_E^{p,n}$ along with 
the matter distribution $G_0(Q)$ of the deuteron gives good agreement with the 
deuteron charge form factor defined by $G_{ch}(Q) = 
G_0(Q)[G_E^p(Q^2) + G_E^n(Q^2) - G_E^p(Q^2) Q^2/(8 m_p^2)]$ 
in \cite{gilmangross} (see Fig. 1). The proton radius corresponding to the dipole 
$G_E^p$ is 0.81 fm and is smaller than that of the free proton radius. However, 
an input of the dipole form of $G_E^p$ reproduces the $ed$ data well as can also 
be found in \cite{arenhovel}. 

\subsection{Deuteron electric potential} 
This potential simply follows from the 
$\mu^-d$ interaction potential in (\ref{potqspace}) by noting that the deuteron electric 
potential should be associated with the Coulomb interaction with form factors but 
cannot depend on the mass of the probe, in this case the muon. Thus, 
\begin{equation}\label{potd} 
V_d(Q) = 4 \pi e \, {G_0(Q) [G_E^p(Q^2) + G_E^n(Q^2)]\over Q^2} \, \biggl ( 1 \, -\, 
{Q^2\over 8 m_N^2} \biggr ) \,, 
\end{equation} 
where $e$ is the positive charge of the deuteron. 
Denoting, 
$G_0(Q)(G_E^p(Q^2) + G_E^n(Q^2)) (1 - Q^2/8m_N^2) = G_d(Q^2)$ so that, 
$V_d(Q) = \, 4 \pi e (G_d(Q^2) / Q^2)$, its Fourier transform is 
the electric potential,  \\
$V_d(r) = 4 \pi e \int e^{i \bm{Q} \cdot \bm{r}} 
(G_d(Q^2)/Q^2)\, d^3Q/(2\pi)^3$, 
the Laplacian of which gives the density $\rho_d(r)$. 

Since we wish to study the sensitivity of the finite size corrections 
corrections to the D-state probability in the deuteron wave function, 
we shall use different parametrizations of the deuteron wave function involving 
about 2 to 7\% of D-state probabilities in order to calculate $G_0(Q)$. 
One choice involves a parametrization of the wave function 
obtained from the Paris NN potential \cite{lacombe} with 
$P_S = \int\, |u(r)|^2 dr = 0.942$ and $P_D = \int\, |w(r)|^2 dr = 0.058$. 
Our second choice is a phenomenological model \cite{berez} 
which uses similar forms as in \cite{lacombe} for parametrizing 
the wave functions but with different parameters, so that the probabilities are
$P_S =  0.983$ and $P_D = 0.017$. Whereas the parameters in \cite{lacombe} 
were fitted to reproduce the numerical values of the Paris wave function, 
those in \cite{berez} were obtained by 
directly fitting the quadrupole moment and deuteron charge form factor data 
with $G_0(Q)$, assuming $G_E^p + G_E^n = 1$. 
In order to test the case with no D-wave component at all, we  perform a calculation 
by normalizing the Paris $u(r)$ in \cite{lacombe} to 1 and not using the D-wave at all. 
We also use an older parametrization \cite{mcgee} 
of the Hamada-Johnston wave functions (once again
having similar forms as in \cite{lacombe} and \cite{berez}) with 
$P_S =  0.93$ and $P_D = 0.7$.
The charge form factor of the deuteron which is extracted 
from scattering experiments seems to be equally well produced 
(considering the error bars and the entire range shown) by 
all choices of the D-state probabilities 
(see Fig. 1). 
\begin{figure}
\includegraphics[width=7cm,height=7cm]{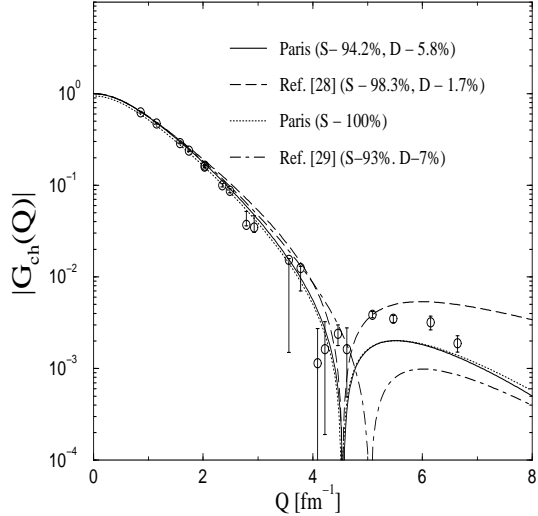}
\caption{\label{fig:eps2} Deuteron charge form factor using different 
D-state probabilities of the deuteron wave function. 
The data is from Ref. \cite{deutffdata}.}
\end{figure}

\subsection{Corrections to the 2S energy levels in muonic deuterium} 
Recent measurements of the 2S-2P transitions in muonic deuterium \cite{pohldeut} 
have shown how precision spectroscopy of atomic energy levels can be used to determine 
the deuteron (and also the proton) radius more accurately than that extracted from 
any scattering experiment. The experiment was based on forming $\mu^-d$ atoms in an 
unstable 2S state and measuring the 2S-2P transitions by pulsed laser spectroscopy. 
The measured value of the 2S-2P Lamb shift is then compared with the theoretical 
calculations involving corrections from Quantum Electrodynamics (QED) and the finite 
size of the deuteron. 
The QED corrections can be calculated very accurately 
\cite{krauth}. 
The finite size corrections (FSC) are incorporated as 
radius ($r_d$) dependent terms. The theoretical value of the Lamb shift 
thus calculated is given by, 
$\Delta E_{LS}^{theo}$ = 228.7766(10) meV + $\Delta E^{TPE}$ - 
6.1103(3) $r_d^2$ meV/fm$^2$, where the second term is a deuteron polarizability 
contribution coming from two-photon exchange and is equal to 
1.7096(200) meV. Comparing $\Delta E_{LS}^{theo}$ with 
the experimentally measured, 
$\Delta E_{LS}^{exp}$ = 202.8785(31)$_{stat}$(14)$_{syst}$ meV, led to the precise 
value of the radius, $r_d$ = 2.12562(13)$_{exp}$(77)$_{theo}$ fm. 
In order to compare the results of the present work with the above precision 
measurements, with the aim of extracting the D-state probability 
in deuteron, we first note that 
the finite size correction (FSC) term, 6.11019 $r_d^2$ meV/fm$^2$ is a sum of 
order $\alpha^4$, $\alpha^5$ and $\alpha^6$ corrections 
given by
6.0731 $r_d^2$, 0.033804 $r_d^2$ and 0.003286 $r_d^2$ respectively.
The order $\alpha^4$ 
part given by 6.0731 $r_d^2$ meV/fm$^2$ will be derived below breifly. 

\subsection{Finite size Coulomb correction at order $\alpha^4$} 
The effect of including the deuteron charge distribution, 
$\rho_d(r)$ in place of the point-like $1/r$ Coulomb potential 
can be incorporated by evaluating the energy correction using first order 
perturbation theory \cite{itzyk}, as, 
\begin{equation}\label{correction2} 
\Delta E_{\rm FS} \, =  \, 
\int |\Psi_{nl}(\bm{r})|^2\, 
\biggl [ e\,V_d(r)\, - \biggl [ -{4\pi \alpha\over r} \biggr ] 
\, \biggr]\, d^3r\, , 
\end{equation}
where, $\Psi_{nl}(\bm{r})$ is the unperturbed atomic wave function and 
$V_d(r)$ is the Fourier transform of the deuteron electric potential 
in Eq. (\ref{potd}). If we now approximate 
$\Psi_{nl}(\bm{r}) \approx \Psi_{nl}(0)$, it is easy to show 
that 
$\Delta E_{\rm FS}$ reduces to \cite{itzyk}, 
\begin{eqnarray}\label{itzykson}
\Delta E_{\rm FS}^0  &=&  {-e \over 6} |\Psi_{nl}(0)|^2 \int d^3r\, 
r^2 \,\nabla^2 V_d(r)\nonumber \\
&=& (2 \pi \alpha/3) |\Psi_{nl}(0)|^2 
\int d^3r \,r^2 \,\rho_d(r) = (2 \pi \alpha/3) |\Psi_{nl}(0)|^2 \langle r^2
\rangle \, ,  
\end{eqnarray}
since, $\nabla^2 V_d(r) = - 4 \pi e \rho_d$. 
For $n = 2$, $l=0$, $(2 \pi \alpha/3) |\Psi_{nl}(0)|^2$ = 6.0731 meV/fm$^2$ 
and the right side of Eq. (\ref{itzykson})
is ${\rm 6.0731}\, \langle r_d^2 \rangle$ as in \cite{pohldeut,krauth}. 
The approximation $\Psi_{nl}(\bm{r}) \approx \Psi_{nl}(0)$ allows us to express the 
FSC in terms of the charge radius and thus 
opens the possibility of determining the charge 
radius of the proton or a nucleus from atomic spectroscopic data which would have 
otherwise been not possible. 
In Table I, we show the tiny difference between the 
calculation of $\Delta E_{FS}$ using (\ref{correction2}) or (\ref{itzykson}).
The table also displays sensitivity of the corrections to the parametrization of the 
deuteron wave function. 
The magnitude of the corrections increases with the lowering of the 
D-state probability in the deuteron wave function. 
It is this sensitivity which leads us to the results shown 
in Table 2 which will be discussed in the next section. 
\begin{table}
\caption{Finite size corrections to the 2S atomic level, $\Delta E_{\rm FS}$ 
(Eq. (\ref{correction2})) and 
$\Delta E_{\rm FS}^0$ (Eq. (\ref{itzykson})).}
\begin{tabular}{|l|l|l|l|l|}
  \hline
\% D-wave  & 7\%\cite{mcgee} & 5.8\%  \cite{lacombe} & 1.7 \%  \cite{berez} & 0 
\cite{lacombe}\\
\hline  
$\Delta E_{\rm FS} $ (meV)& 26.2 & 26.72 & 27.03 & 27.57 \\
$\Delta E_{\rm FS}^0$ (meV)& 26.53 & 27.01 & 27.35 & 27.87 \\
\hline 
\end{tabular}
\end{table}
Note that even though there exists a tiny difference in the values of 
$\Delta E_{\rm FS}$ and $\Delta E_{\rm FS}^0$ in Table 1, for the comparison of 
the radius evaluated from $r_d = \Delta E_{\rm FS}^0/6.0731$ with $r_d^{\rm exp}$
which has been fitted to data using a similar formula, this difference does not matter. 

\section{Deuteron charge radius and D-state probability}
The  electric potential $V_d(Q)$ in (\ref{potd}) can also be expressed as 
$V_d(Q) =  4 \pi e G_{ch}(Q)/Q^2$ with, 
$G_{ch}(Q) = 
G_0(Q) [G_E^p(Q^2) + G_E^n(Q^2)] [1 \, -\, 
(Q^2/8 m_N^2) ]$ being 
the Fourier transform of $\rho_d(r)$, so that using standard formulae for the 
expressions connecting radii and form factors \cite{weproton}, we obtain, 
$r_d^2 = r_p^2 + r_n^2 + (3 / 4 m_p^2) + (1/4) \, 
\int_0^{\infty} \, [ |u(r)|^2 + |w(r)|^2 ] \, r^2 dr$ , 
where, the last term is the matter radius $r_m^2 = -6 (dG_0/dQ^2)|_{Q^2=0}$. 
Thus, for a given parametrization of $G_E^p$ and $G_E^n$ which 
reproduce the data on $G_{ch}(Q)$ as defined above well (see Fig. 1), 
$r_d$ can be seen to depend on the deuteron wave function $w(r)$. 
By choosing a certain $w(r)$, we choose also a certain $P_D$, since 
$P_D = \int |w(r)|^2 dr$. Knowing the values of 
$\Delta E_{\rm FS}^0$ (see second line of Table 1), the radius can be determined 
from Eq. (\ref{itzykson}), namely, 
$\Delta E_{\rm FS}^0 = 6.0731 \, r_d^2$. Since the fits in \cite{pohldeut} assume 
a similar form of the $\alpha^4$ FSC, it is appropriate to compare this $r_d$ 
with the fitted value of $r_d^{\rm exp} = 2.12562(78)$ fm in \cite{pohldeut}. 
\begin{table}
\caption{The estimated values of the radius 
for different D-state probabilities.}
\begin{tabular}{|l|l|l|l|}
  \hline
 Model & &\% $P_D$ & $r_d$ (fm)\\ 
\hline  
       & EGM N3LO & 3.28 & 2.1315 \\
Chiral & EMN N4LO & 4.1 & 2.1277 \\
Ref. \cite{chiralNN}& J\"ulich N4LO & 4.29 & 2.1268 \\
       & EM N3LO & 4.51 & 2.1296\\
\hline
               & CD-Bonn & 4.85 & 2.1212 \\
High precision & NjmNR & 5.635 & 2.1222 \\
 Ref. \cite{highNN}& NjmR & 5.664 & 2.1226\\
               & Reid93 & 5.699 & 2.1236 \\
               & AV18 & 5.75 & 2.1221 \\
               & Paris \cite{lacombe} & 5.8 & 2.126 \\
\hline
               & TRS & 5.92& 2.1297 \\
Traditional    & RSC & 6.47& 2.1127 \\
 Ref. \cite{tradNN} & RHC & 6.497& 2.1156 \\
               & HW  & 6.953& 2.1223 \\
               & McGee \cite{mcgee}& 7& 2.107\\
\hline
Phenomenological \cite{berez}& & 1.7& 2.14 \\
\hline
\end{tabular}
\end{table}
\begin{figure}[h]
\includegraphics[width=9cm,height=9cm]{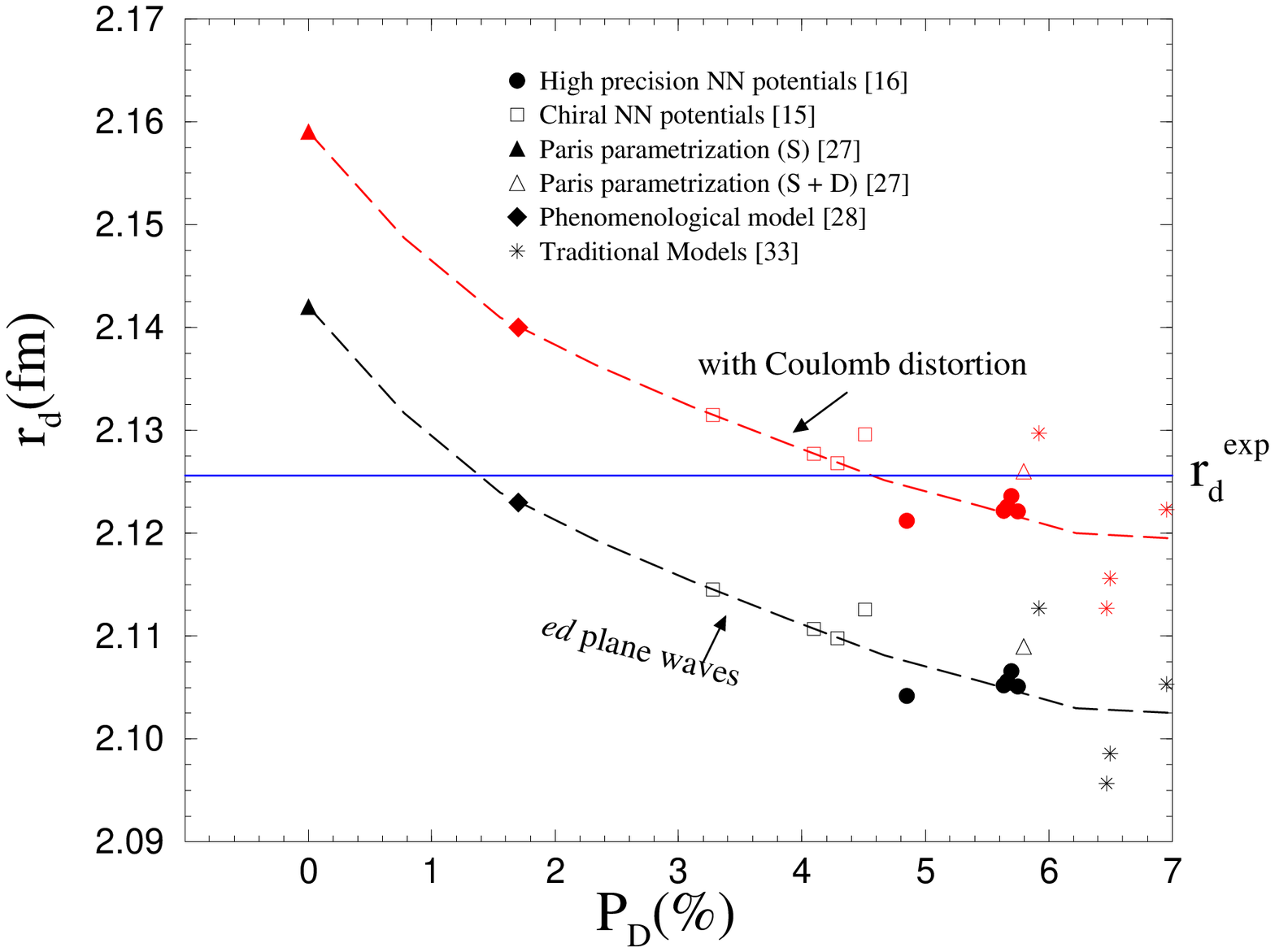}
\caption{Charge radius of the deuteron as a function of 
the D-state probability in the deuteron wave function 
evaluated using different models. 
The dashed lines are drawn to guide the eye with the red (black) 
one representing 
the numbers obtained with (without) the effects of Coulomb distortion included. 
}
\end{figure}
In studies of electron-deuteron scattering as in Ref. \cite{gilmangross}, 
data have been interpreted in terms of the plane wave Born 
approximation (PWBA). 
However, the effects of including distorted $ed$ waves can become 
important for comparisons with precise data. 
Noting that the Coulomb distortion \cite{sicktrautman} 
changes the deuteron radius by 0.017 fm, 
the authors in \cite{gilmangross} suggest 
an adjustment of the deuteron charge form factor by an amount 
$\delta G_C = -0.003\,+0.104\, Q^2$ which decreases the form factor at small 
$Q^2$ and increases the value of the radius. The results presented 
in Table 1, however,  do not take these effects 
into account since it is not  
appropriate to evaluate (\ref{correction2}) (which involves $V_d(r)$ obtained 
from a Fourier integral over all momenta) using the form factor corrected 
only at low $Q^2$. The correction at low $Q^2$ 
introduces a disagreement with data at large $Q^2$ as 
shown in \cite{gilmangross}. The above correction is however important for 
the calculation of the radius defined by the derivative of the form factor 
at $Q^2$ = 0 and hence  
in Table 2, we present the deuteron charge radius, $r_d$, with  
the Coulomb distortion correction of 0.017 fm as in \cite{sicktrautman} 
for different choices of 
the nucleon-nucleon (NN) potentials. Fig. 2 displays the same with and without 
the Coulomb distortion included.   
From the figure we observe a general trend of increasing $P_D$ for smaller 
radii. However, the results are model dependent with the chiral models 
indicating a value of $P_D$ = 4.3 and the high precision NN models a value 
around 5.7 leading to a good agreement with the experimental 
$r_d = 2.12562(78)$ fm \cite{pohldeut}. The choice of the proton and neutron 
form factor parametrization (which affects the values of $r_p$ and $r_n$ 
entering in   
$r_d^2 = r_p^2 + r_n^2 + (3 / 4 m_p^2) + (1/4) \, 
\int_0^{\infty} \, [ |u(r)|^2 + |w(r)|^2 ] \, r^2 dr$), can add a small 
uncertainty to the values deduced in Table 2. The magnitude of these  
uncertainties using different parametrizations for $G_E^p$ and $G_E^n$ 
which reproduce the deuteron charge form factor, 
$G_{ch}(Q) = G_0(Q) [G_E^p(Q^2) + G_E^n(Q^2)] [1 \, -\, 
(Q^2/8 m_N^2) ]$ equally well, remains to be investigated in future.

\section{Finite size Coulomb plus Darwin corrections at order $\alpha^4$}
For completeness, we also calculate the FSC with the Darwin terms in (\ref{potqspace})
within the few body formalism. 
The Fourier transform of the muon-deuteron interaction potential, 
$V_{\mu^- d} (Q)$, can be done numerically to obtain 
the potential in coordinate space which can then be used to evaluate the energy 
correction using first order time independent perturbation theory as, 
\begin{equation}\label{correction}
\Delta E = \int |\Psi_{nl}(\bm{r})|^2\, 
\biggl [ \,V_{\mu^- d} (r)\, - V_{\mu^-d}(r)^{point} 
\, \biggr]\, d^3r\,  
\end{equation} 
where, $\Psi_{nl}(\bm{r})$ is the unperturbed atomic wave function. 
Note that we have subtracted 
the point-like contribution $-\alpha/r$ as well as the point-like Darwin terms  
from $V_{\mu^- d} (r)$ so that the quantity 
in square brackets is the perturbative potential only due to deuteron structure. 
In Table 3, we list the finite size corrections (FSC) 
(to the Coulomb and Darwin terms) of order $\alpha^4$ in muonic 
deuterium using Eq. (\ref{correction}) and 
different percentages of the D-state  
probabilities in the deuteron wave function for the energy levels 
with $l=0$ and $n =1, 2$. 
Since the numbers in Table 3 are not very different from those in Table 1
(compare the first line in Table 1 with the second line in Table 3), 
one can say that the FSC due to the Darwin terms are in general very small.
\begin{table}
\caption{Finite size corrections $\Delta E$ in meV 
(Eq. (\ref{correction})), 
to the 1S and 2S atomic levels in $\mu^-d$, 
for different D-state probabilities of the deuteron wave function.}
\begin{tabular}{|l|l|l|l|l|}
  \hline
  &7\% \cite{mcgee} & 5.8\% \cite{lacombe} & 1.7 \%\cite{berez} & 0 \cite{lacombe}\\
\hline  
1S & 206.28 & 210.42 & 213.17  & 217.19 \\
\hline 
2S& 25.78 & 26.31 & 26.65  & 27.15 \\
\hline
\end{tabular}
\end{table}
Note that Eqs (\ref{correction}) and (\ref{correction2}) are different in the sense 
that (i) $V_{\mu^-d}$ in (\ref{correction}) contains the additional muon Darwin 
term as compared to $e \,V_d$ and (ii) 
whereas (\ref{correction}) subtracts the point-like potential, 
$V_{\mu^-d}(r)^{point}$, which contains the point-like Coulomb term, 
$-4 \pi \alpha/r$ and 
two point-like Darwin terms $\delta^3(\bm {r})/8M_N^2$ and 
$\delta^3(\bm {r})/8M_{\mu}^2$, Eq. (\ref{correction2}) subtracts 
only the point-like Coulomb, $- 4 \pi \alpha/r$. 

To summarize, the leading order nuclear structure corrections in muonic deuterium 
have been evaluated within a few body formalism which reveals the dependence 
of the correction on the model of the deuteron 
wave function. Since scattering data do not have the 
high precision achieved by the muonic atom spectroscopy data, the 
deuteron charge form factor can be reproduced equally well 
(within error bars) by all the parametrizations of the deuteron 
wave function used, irrespective of 
the percentage of $P_D$ in them. However, we notice that a 
comparison of the radius evaluated using these parametrizations with 
the precise radius value extracted from $\mu^- d$ spectroscopy provides 
a complementary tool to determine $P_D$. Though there do exist 
model dependent uncertainties in $P_D$ (see Table 2 and Fig. 2), there 
seems to be a general trend of increasing values of $P_D$ for smaller 
$r_d$.      
The few body formalism presented here can also be used to evaluate 
the nuclear structure corrections in muonic helium atoms which are 
expected to be studied in future.  
\\
\\
{\bf Acknowledgments}\\
One of the authors (D. B. F.) thanks the  administrative  department  of  science,  
technology  and  innovation of Colombia (COLCIENCIAS) 
and the Faculty of Science, Universidad de los Andes for  the financial support 
provided. He is also grateful to IFIC, University of Valencia and Prof. Vicente 
Vacas for their kind hospitality and useful discussions. 

\end{document}